# Enabling decentral collaborative innovation processes - a web based real time collaboration platform

Matthias Joiko[1], Florian Kohnen[1], Kevin Lapinski[1], Houda Moudrik[1], Irawan Nurhas[2], Florian Paproth[1] and Jan M. Pawlowski[2]

[1] Hochschule Ruhr West, Institute of Computer Science, Bottrop, Germany
{matthias.joiko,florian.kohnen,kevin.lapinski,houda.moudrik, florian.paproth}@stud.hs-ruhrwest.de

[2] Hochschule Ruhr West, Institute of Computer Science, Bottrop, Germany
{irawan.nurhas,jan.pawlowski}@hs-ruhrwest.de

**Abstract.** The main goal of this paper is to define a collaborative innovation process as well as a supporting tool. It is motivated through the increasing competition on global markets and the resultant propagation of decentralized projects with a high demand of innovative collaboration in global contexts. It bases on a project accomplished by the author group. A detailed literature review and the action design research methodology of the project led to an enhanced process model for decentral collaborative innovation processes and a basic realization of a browser based real time tool to enable these processes. The initial evaluation in a practical distributed setting has shown that the created tool is a useful way to support collaborative innovation processes.

## 1    Introduction

Today's companies must cope with rapid technological dynamics and the growing competition on global markets. To master the upcoming challenges steady innovation of processes and products is mandatory. Enabling steady innovations makes it necessary to implement collaborative innovation process models [1, 2]. These process models must be well designed to make the most out of the advantages of collaborative working under consideration of the risks of decentral projects. Well-engineered innovation process models are available [2–4]. What misses in the established models is the consideration of decentral collaborative real time working groups in distributed / global contexts. This paper works out the core processes of established innovation process models to refine them and achieve adaption to these scenarios. For instance, consideration of globalization and localization regarding cultural differences is fundamental.

Decentral collaboration depends on tools which enable communication and interaction between the collaboration group members in real time [5]. Previous research comes to the result, that the market offers a variety of tools which support collaboration, partially also in real time, but not even one provides a simple, holistic approach especially for innovation processes. Also, no fitting tool is available as open source software.

The main purpose of this paper is to outline a collaborative innovation process which is enhanced by web based software. The process model, based on an action design research method, is based on established innovation process models [6, 7] to provide structured stages from idea creation to market release. Finally, the paper provides a platform prototype tool based on latest know-how about collaborative innovation processes in a holistic approach.

## 2 State of the art

In the following section, we analyze innovation process models as well as existing web based tools which can support collaborative processes.

### 1.1 Overview of Innovation Process Models

In general, an innovation process is defined as systematic realization of existing and newly created product ideas [8]. The realization covers all steps from idea creation, prototyping up to the final market release. An innovation process aims for timely realization of a product idea to a market stable and releasable product, but is dependent to available capabilities of an organization [9]. This kind of systematic procedure requires special attention on flexibility to be able to react on late or rushed modifications. Several models have been used for the past decades as a reference, such as Schumpeter's or Roger's innovation models [10–12]. Those innovation process models have evolved in recent decades from simple linear and sequential models to increasingly complex models embodying a diverse range of inter and intra stakeholder and process interaction. Our starting point is the first and second-generation models, explaining innovation as either being pulled by market needs, or pushed by technology and science [12]. An obvious disadvantage of their linear process is the distortion of the real innovation process and its formalization and it is impossible to transmit high-quality communication between the components of the innovation process [13]. A further model is a coupling model that recognizes the influence of technological capabilities and market needs within the framework of the innovating firm. However, most of the innovation process models in literature involve a pattern of the following steps or stages: (a) idea generation and identification, (b) concept development, (c) concept evaluation and selection, (d) development, and (e) implementation. Furthermore, all Innovation process model can either be a market pull or technology push, or even a combination [13].

### 1.2 Collaborative Innovation Process

Collaborative innovation processes are innovation processes, that are geographically distributed and can be realized over frontiers. Innovative new products are developed with shared resources like knowledge or industrial capabilities. Collaborative

1532

innovation requires a more complex coordination and needs a structured way to communicate. A successful innovation process supports real-time collaboration and communication by multidisciplinary teams across different countries and time zones. In order to create an efficient collaborative innovation process, an innovation workflow, with supporting software tools and artefacts, has to be defined.

Pallot et al [14] identified barriers and specifically difficulties to gain trust between collaboration partners but shows that most realized collaborations lead to faster technological process. Collaborative Innovation is based on many success factors, amongst them:

- Strong focus on modelling the value needs of the different stakeholders
- Real Time communication, which stakeholders can directly and openly exchange ideas and information.
- Use of purposive inflows and outflows of knowledge to accelerate internal innovation and expand the markets for external use of innovation
- More collaborative, and more engaging with a wider variety of participants.
- Users are expected to share their knowledge freely within the community because as users they benefit directly from innovation
- Users are a powerful source of innovation in the initial stages of a new product
- Use of internal / external ideas as well as internal / external paths to market

A Collaborative Innovation model is needed incorporating those aspects. The innovation pyramid by Zeidner & Woods [15] meets these criteria because it opens the innovation cycle to a variety of actors and taps into innovation resources across borders, overcomes cultural restrictions.

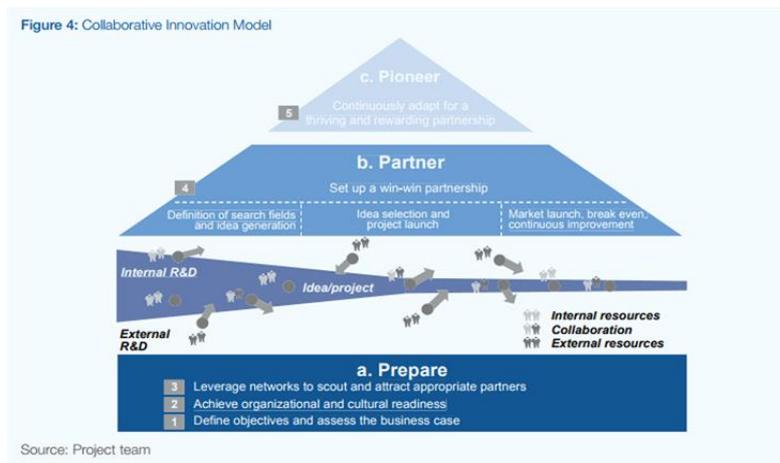

**Figure 1 - Collaborative Innovation Pyramid [14]**

The preparation layer in this Pyramid lays the critically-important and often overlooked foundation for collaboration, and involves defining objectives and connecting with the right potential partners [16]. Furthermore, the partnering layer focuses on negotiating and tailoring the projects with partners to ensure that the benefits, risks and governance

1533

aspects are adequately defined. Finally, the pioneer layer ensures that partnerships adapt and thrive for the mutual and sustained benefit of all parties as they are executed and as the context changes. Due to the fit of detail, we choose this model as a base for our extended process model, described in detail in section 4.

## 2    Methodology and used tools

The underlying approach of the project is Action Design Research [17]. The stages implemented in this project are the definition of the leading scientific interrogation, developing and evaluation, reflection and formalization of lessons learned. To achieve an appropriate answer to the central question the project follows practice-inspired research to gain a theory-ingrained artefact. The developing and evaluation step ensures the technological progress and the ongoing sharpening of the final good by considering expert suggestions in weekly meetings. The reflection stage of the action depends on the input parameters of the two stages described above as the team must reflect the developing process and the evaluation feedback in relation to the central research question. Thus, ADR methodology allows a combination of creating, intervening and evaluating during a focused research process [17]. This leads to early and fast alpha versions and continuous improvements without losing sight of the big picture. To enable these processes for the design of the enhanced decentral collaboration process model and the development of the final web based software application the team uses mainly two tools: Git to support versioning and general administration of the developed source code and Signavio to collaboratively design processes [18, 19, 20].

## 3    Innovation Process Model

Based on the concepts shown above, we refined an innovation process incorporating a focus on networking and collaboration. The innovation environment changes through networking and collaboration from simple linear models to more complex integrated collaboration models. In this new networked paradigm, it's possible to combine linear and coupling processes depending on the requirements. -To build our collaborative innovation process model, three connected level of abstraction were created. The first level of the process model is a value chain diagram. It provides a quick overview of the phases in the innovation management process. The figure below shows these phases of the developed innovation process by using a collapsed process that contains a sub-process und can be linked to the next element of another diagram.

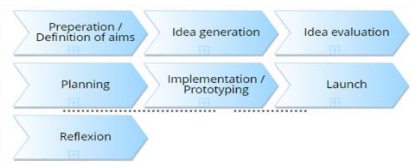

Figure 3: Collaborative Innovation Process Model - Value chain

The first step 'Preparation' in our process is to identify problems, needs and expectations for the project. And as a result, new ideas are created during idea generation stage. However, inspiration for a new idea can originate from an improvement of an existing idea, or something from scratch. Once the relevant ideas and references are provided, their accuracies will be verified, evaluated and the potentials will be measured by evaluating benefits and problems in "Idea evaluation" stage. The planning step will start once the ideas and information have been collated and verified. After that, in prototyping stage begins initial prototype. At this stage, the initial model based on the ideas provided will be created and will be shared with the different stakeholders for comment and review. After completion of all iterative versions of product, a second meeting would be scheduled with Client and Partner to review the 3D model and the last product. At least the Marketing and reflection stage are the stage which include production, market launch and penetration in national or international level.

To create the second level of our innovative process model we use Value Chain Diagrams. The following figure shows an example, the idea generation stage.

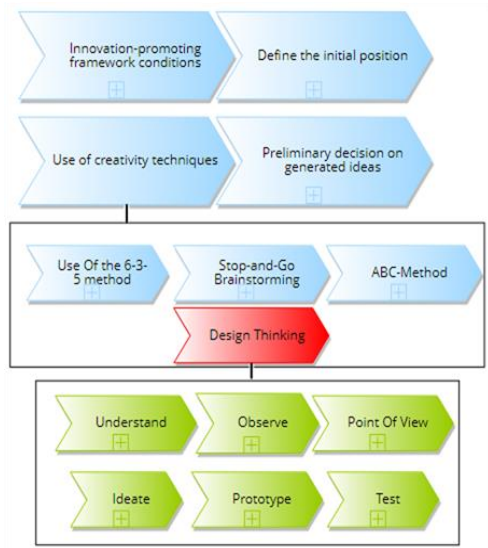

**Figure 4: Idea generation stage - Value chain**

This figure gives a short insight on the different stages of successful ideation and idea generation techniques. For example, the idea generation techniques identified are briefly introduced as follows:
- Design Thinking [21]
- 6-3-5 Method [22]
- ABC-Method [23]
- Stop-and-Go Brainstorming [24]

In the third level of our integrated innovative process, we modeled detailed, event-driven process chains. The following figure shows an example of the developed process chain diagrams. It describes the collaborative prototyping process within the "Implementation/Prototyping" stage of the top-level value chain.

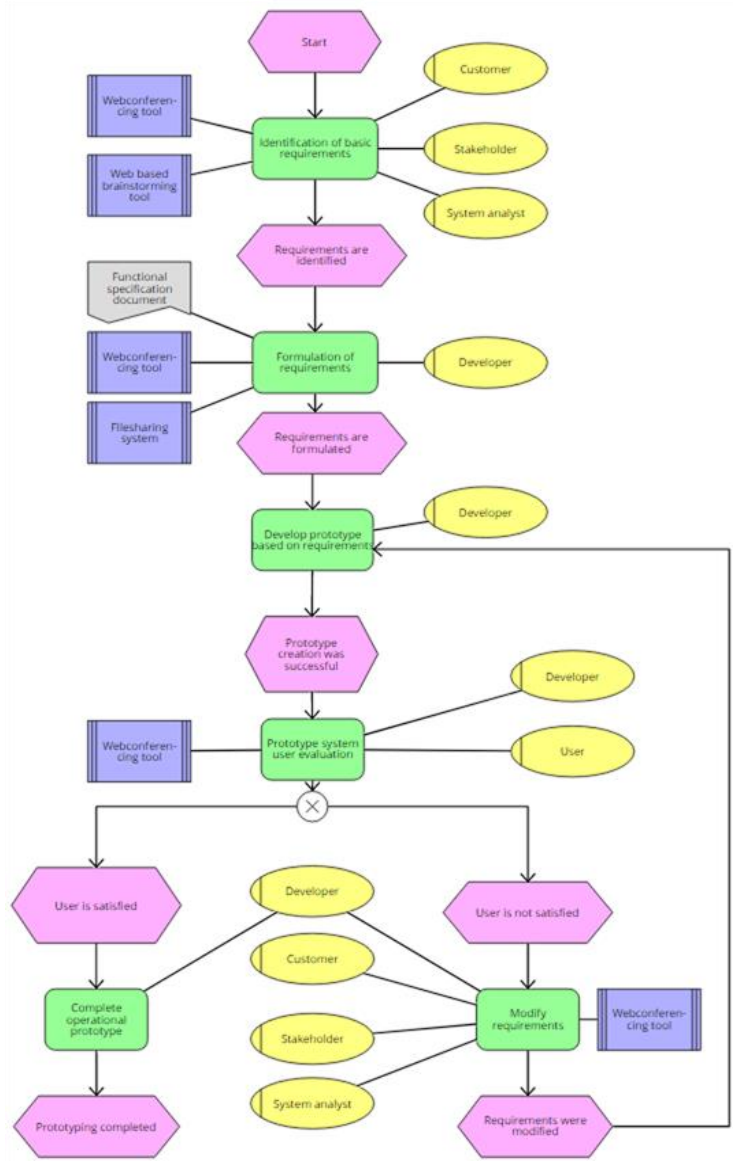

**Figure 5: Collaborative Prototyping - EPC Model**

In addition, it should be said that there are several ways of approaching collaboration innovation processes. In our case, we had to integrate the different Stakeholders in each step of our process. Ideas are realized in cooperation with customers, users and partners either via virtual networks or real meetings. Also communication and sharing knowledge in different steps are specific for the Collaborative Innovation Process. As a summary, the model can be used as a reference for future innovation processes in different settings.

## 4       Collaborative Innovation Support Tool

The basic idea of our tool is that it should support a range of core processes defined by the innovation process model described above. Furthermore, it should be a system platform independent browser based solution. The tool is based on a previous research on open innovation [25]. The work provided requirements and recommendations for global authoring platforms under consideration of various barriers as "lack of technical skills, different types of devices and systems as well as the cultural differences in cross-border collaboration' [25]. It also provided a web based platform implementation, which is the foundation of the development described in the following paragraphs of this chapter. The platform is characterized by real time synchronized presentations, where members of the collaboration teams can create, edit and delete slides. Slides can contain text, images and survey elements.
To enhance the platform focusing on collaborative innovation, we decided to implement a new type of presentation slides including "sticky notes". As this raises the complexity of the presentation structure significantly, we had to implement an enhanced, more intuitive site navigation. This navigation includes the usage of linked block titles and predefined jumping points. As the common understanding of "sticky notes" includes, the notes must be enabled for free and direct positioning within the borders of the containing slide including overlapping. Obviously, we implemented highlighting notes with color and linking notes among each other with simple lines, like the advantages of a physical, offline whiteboard also provide. To integrate single sticky notes into the site navigation it is possible to link notes, so you can jump to them wherever a reference link to it appears. Vice versa it is also possible to link presentation block titles within a note. Also, external hyperlinks and further common attachments can be applied to the sticky notes. The sticky note approach is a very good foundation to enable a lot of core process chains described by the collaborative innovation process model. For this purpose, we decided to provide five sticky note background design templates as built-in feature of the platform. Specially to accelerate the idea creation, evaluation and reflection stages, templates for design thinking methods, free sticky walls as well as SWOT analysis are available. Also, we provide general Kanban and Scum templates to enable these project management methods.

The following figures show an example of usage for the tool based on a real collaborative innovation project, initiated by the HRW FabLab [26]. The project aimed at creating a sensor-based dog harness allowing to analyze dogs' emotions including hardware as well as an app for monitoring. The project was realized involving different stakeholders distributed across the globe. The first figure shows exemplary how the SWOT template of the sticky note platform feature can be used to evaluate the "remote dog harness" idea. It also shows how the sticky note linking mechanism works.

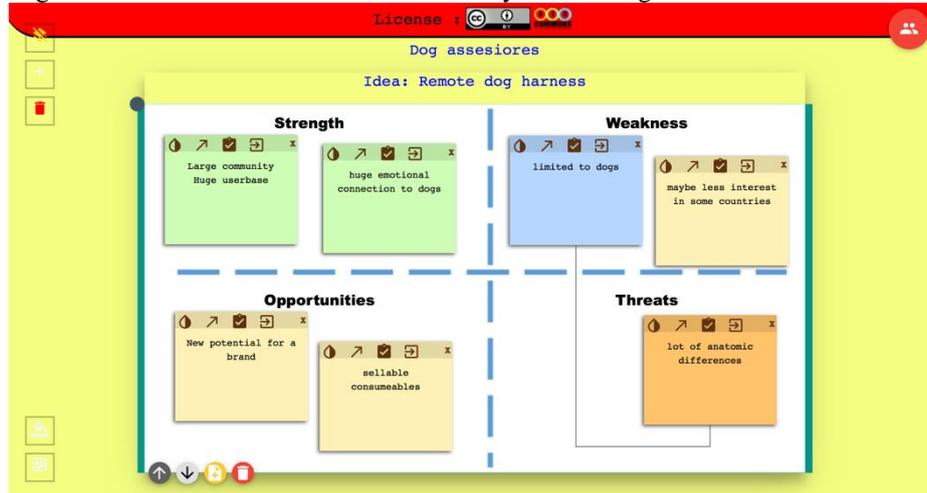

**Figure 6 – Platform screenshot – Idea evaluation SWOT**

The second figure shows one of the possibilities to manage upcoming tasks while the implementation phase. A Kanban board is used in this example. It illustrates a highlighted sticky note and how transitional task stages can be shown on the board.

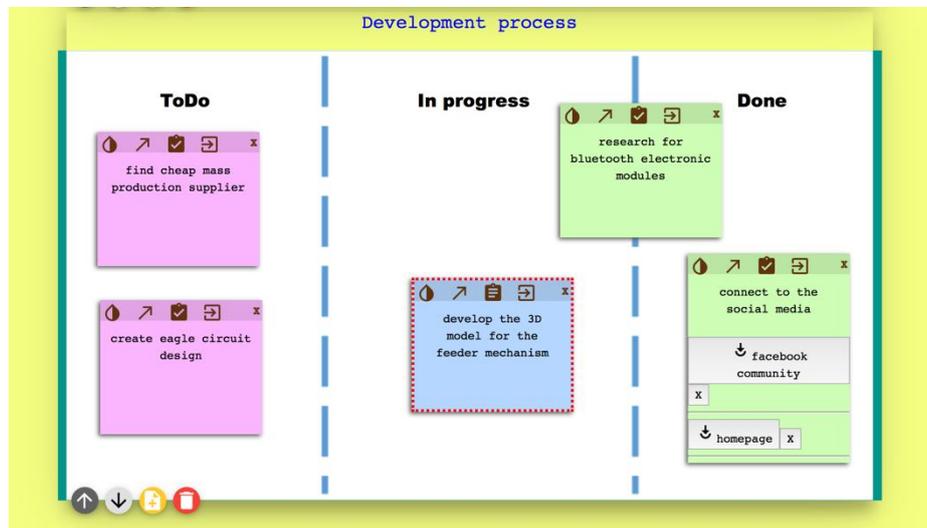

**Figure 7 – Platform screenshot – Manage tasks**

The part of the software which implements the sticky notes and the templates uses the Node.js ® event driven JavaScript runtime built and Webstrates, which is a research prototype for collaborative editing websites and client synchronisation through HTML DOM manipulations [27, 28]. Sticky note linking is implemented on basis of the jsplumb toolkit [29]. As a matter of course, internationalization and localization (i18n and l10n) software engineering methods are implemented in every developed software part during this research project [30].

The sample case involved different users which were coordinated by the initiating stakeholder. All phases were realized collaboratively utilizing the tool as the main collaboration and feedback support.

## 5    Proof of concept

This chapter contains two steps to proof the developed concept and tool on top of observing experiences in the initial phase as described in the last chapter. We use a scenario-based evaluation as a first step. The second step was realized by conducting a live tool test and survey with a focus group of researchers and practitioners.

### 5.1    Scenario

To proof the developed concept, we consider a scenario in which a German company works cooperative with a national park in Finland. Together they try to generate and evaluate new ideas for a touristic attraction in the national park. Due to the distributed workplaces, the employees work collaborative via the online platform OERauthors [30]. In the beginning all members get access to the project, stored on the online platform with a shared unique Project-URL. After connecting to it with a browser, the user can start and his following interactions are sent immediately to all members. Since the whole project is localized, any user can decide which system language he wants to work with. The project manager opens a brainstorm-template in which all ideas can be written down on virtual sticky notes. Any member can create, modify and delete notes. Since the number of sticky notes can vary, there are different types of templates to select. An overview- and a detailed perspective, which can handle more notes without becoming confusing. Helping to categorize the generated ideas in the brainstorming process, it is possible to set different background colors for single notes. After a sufficient number of ideas has been gathered during brainstorming, these need to be evaluated with help of the next template including a SWOT analysis. The ideas to be assessed are written down on notes again and positioned in the corresponding quadrant of the SWOT-template. To discuss various decisions, the members communicate over an integrated chat function. To set the focus of discussion they can highlight sticky notes.

The next step in the innovation process deals with the planning phase, which is also supported by templates. The project manager has the choice between Kanban or Scrum. Just to illustrate the status of all necessary to-do's he chooses the Scrum-template and creates the topic notes. To make the document scalable, the Project manager links notes to templates which describe the subject in more detail. With help of the implemented navigation, the members can jump to various templates directly. Media files like videos and pdf-documents are stored on additional cloud server but the links to open them are also saved in notes placed on the template.
This short scenario has been used to show that both templates and tools are useful to support different phases of the innovation process and can be realized in distributed settings.

### 5.2 Survey

To validate the concept and tool in detail, we made a live test of the developed tool with a group of four experts from science and practice. Subsequently we conducted a system usability scale based survey with the expert group [31]. Additionally, the survey contained interview questions with the possibility to add free text comments. Based on the survey, the system received the score on the average system usability scale of 83,75. Additional free text comments showed positive feedback and a high system quality. The gathered results lead to the conclusion that the developed tool enables the acquired collaborative innovation process model in an intuitive way. Not a single free text comment contained a doubt, question or discredit regarding the developed concept. Most of the comments were suggestions for further general user interface features. These perceptions are the focus of the outlook chapter, which follows at the end of this paper.

## 6   Conclusion & Outlook

The project summarized in this paper was motivated through the increasing competition on global markets and the resultant propagation of decentralized projects with a high demand of decentral innovative collaboration in global contexts. A detailed literature review and the action design research methodology iterations led to progressive reshaping of the developed concept and tool until they achieved the predefined goals.
This paper consolidates core processes of enhanced collaborative innovation process models with state of the art web based software engineering, as it contains a new innovation process model and an enhanced software tool for the integration of the processes into an open authoring platform.
As mentioned in proof of concept, the conducted survey and especially the free text comments brought new perceptions for the future development of the open authoring platform the project was based on and for the tool developed during the project. For instance, a sample of the survey incites to combine the sticky note specific controls with the navigation control elements of the platform in one single control bar. Also, the navigation control design could be more like a classical table of contents in future versions of the platform. Further improvement could be achieved by implementing several sharing functionalities as export to .PDF documents or the possibility to share created projects directly on social media. To promote these features, larger and more scalable sticky note templates could be implemented in future also.

In summary it can be said, that this research project brings a new approach of decentral collaborative innovation and an intuitive web based tool, which both can be very useful in various future projects.